\documentclass{PoS}

\title{The high density equation of state: constraints 
from accelerators and astrophysics}

\ShortTitle{The high density equation of state}

\author{\speaker{Christian Fuchs}\thanks{A footnote may follow.}\\
Institute of Theoretical Physics, University of T\"ubingen, Germany\\
        E-mail: \email{christian.fuchs@uni-tuebingen.de}}

\abstract{
The nuclear equation of state (EoS) at high densities and/or 
extreme isospin is one of the longstanding problems of nuclear 
physics. In the last  years 
substantial progress has been 
made to constrain the EoS both, from the astrophysical side and from 
accelerator based experiments. Heavy ion 
experiments support a soft EoS at moderate densities while 
the possible existence of high mass neutron star observations 
favors a stiff EoS. 
Ab initio calculations for the nuclear many-body problem make 
predictions for the density and isospin dependence of the 
EoS far away from the saturation point.  
Both, the constraints from  astrophysics and accelerator based experiments 
 are shown to be in agreement 
with the predictions from many-body theory.}

\FullConference{Critical Point and Onset of Deconfinement
          4th International Workshop\\
		 July 9-13 2007\\
		 GSI Darmstadt,Germany}

\begin{document}
\section{Introduction}
The isospin dependence of 
the nuclear forces which at present is only little 
constrained by data will be explored by the forthcoming radioactive beam 
facilities at FAIR/GSI, SPIRAL2/GANIL and RIA. 
Since the knowledge of the  
nuclear equation-of-state (EoS) at supra-normal densities and extreme 
isospin is essential for  
our understanding of the nuclear forces as well as for astrophysical  
purposes, the determination of the EoS was already one of the primary  
goals when first relativistic heavy ion beams started to operate. 
A major result of the SIS100 program at the GSI 
is the observation of a soft EoS for symmetric matter in the 
explored density range up to 2-3 times saturation density. These
accelerator based experiments are complemented by astrophysical 
observations. 

In particular the stabilization of high mass neutron 
stars requires a stiff EoS at high densities. There exist several 
observations pointing in this direction, e.g. the large radius of $R > 12$ km for the 
isolated neutron star RX J1856.5-3754 (shorthand: RX J1856) 
\cite{Trumper:2003we}. 
Measurements of high masses are also reported for compact stars in low-mass
X-ray binaries (LMXBs) as $M=2.0\pm 0.1~M_\odot$ for the compact object in 4U 1636-536
\cite{Barret:2005wd}.    
For another LMXB, EXO 0748-676, 
constraints for the mass $M\ge 2.10\pm 0.28~M_\odot$
{\it and} the radius $R \ge 13.8 \pm 0.18$ km 
have been reported \cite{Ozel:2006km}.
Unfortunately, one of the most prominent high mass neutron star 
candidates, the J0751+1807 binary pulsar with an originally reported 
mass of $M=2.1\pm 0.2~M_\odot$ \cite{NiSp05} has been revisited and 
corrected down to $M=1.26~M_\odot$ \cite{Nice07}. However, 
very recently an extrodinary high 
value of  $M= 2.74\pm 0.21~M_\odot$ $(1\sigma$) has been reported for 
the millisecond pulsar PSR J1748-2021B \cite{Freire:2007jd}.

Contrary to a naive expectation, high mass neutron stars 
do, however, not stand in contradiction with the observations from 
heavy ion reactions, see e.g. \cite{Fuchs:2007vt,Sagert:2007nt}. 
Moreover,  we are in the fortunate situation that 
ab initio calculations of the nuclear many-body problem predict a density and 
isospin behavior of the EoS which is in agreement with both observations.

Hence the present contribution starts with short survey on the predictions 
from many-body theory, turns then to heavy ion reactions and discusses 
finally the application to neutron stars.
\section{The EoS from ab initio calculations} 
In  {\it ab initio} calculations based on many-body techniques one  
derives the EoS from first principles, i.e. treating  
short-range and  many-body correlations explicitly. This allows to 
make prediction for the high density behavior, at least in a range where 
hadrons are still the relevant degrees of freedom. A typical  
example for a successful  many-body approach is Brueckner theory (for 
a recent review see \cite{baldo07}). In the following we consider 
non-relativistic Brueckner and  variational  calculations \cite{akmal98} as 
well as relativistic Brueckner calculations \cite{boelting99,dalen04,krastev06}. 
It is a 
well known fact that non-relativistic approaches require the inclusion 
of - in net repulsive - three-body forces in order to obtain reasonable 
saturation properties. In relativistic treatments part of such diagrams, 
e.g.  virtual excitations of nucleon-antinucleon pairs are already effectively 
included. 
\begin{figure}[h] 
\centerline{
\includegraphics[width=0.8\textwidth]{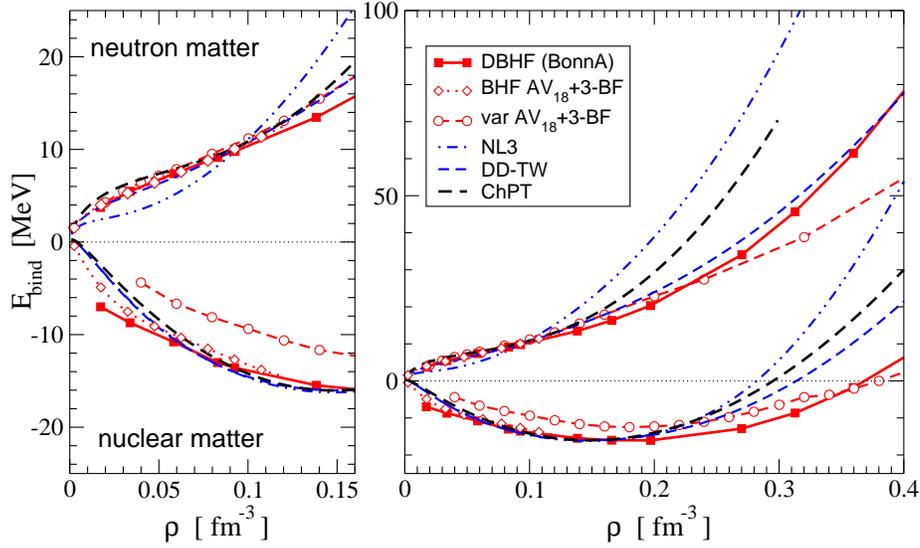}}
\caption{EoS in nuclear matter and neutron matter.  
BHF/DBHF and variational calculations are compared to  
phenomenological density functionals (NL3, DD-TW) and  
ChPT+corr.. The left panel zooms the low density range. 
The Figure is taken from Ref. \protect\cite{WCI}.  
} 
\label{nmeos_fig} 
\end{figure} 
Fig. \ref{nmeos_fig} compares now the predictions for nuclear and neutron  
matter from microscopic  
many-body calculations -- DBHF \cite{dalen04}  and the 'best' 
variational calculation with 3-BFs and boost corrections  
\cite{akmal98} -- to phenomenological approaches (NL3 and DD-TW 
from \cite{typel99}) and an approach based on chiral pion-nucleon dynamics 
\cite{finelli} (ChPT+corr.). As expected the phenomenological functionals  
agree well at and below saturation density where they are constrained  
by finite nuclei, but start to deviate substantially at supra-normal  
densities. In neutron matter the situation is even worse since  
the isospin dependence of the phenomenological functionals is less constrained.  
The predictive power of such density functionals  at supra-normal  
densities is restricted.   {\it Ab initio}  
calculations predict throughout a soft EoS in the density range  
relevant for heavy ion reactions at intermediate and low  
energies, i.e. up to about 3 $\rho_0$.  
Since the $nn$ scattering length is large, neutron matter 
at subnuclear densities is less  
model dependent. The microscopic 
calculations (BHF/DBHF, 
variational) agree well and results are consistent with 
 'exact' Quantum-Monte-Carlo calculations \cite{carlson03}. 
\begin{figure}[h] 
\centerline{
\includegraphics[width=0.8\textwidth]{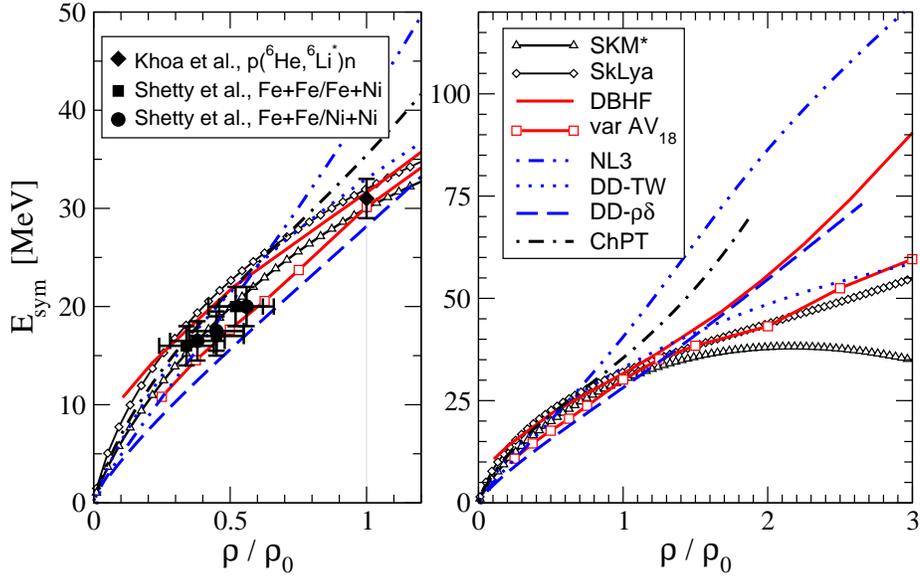}} 
\caption{Symmetry energy as a function of density as predicted by different  
models. The left panel shows the low density region while the right  
panel displays the high density range. Data are taken from \protect\cite{Khoa06} 
and \protect\cite{shetty07}.
} 
\label{esym_fig} 
\end{figure} 

Fig. \ref{esym_fig} compares the symmetry energy predicted from  
the DBHF and variational calculations to that of the  
empirical density functionals already shown in Fig. \ref{nmeos_fig}.  
In addition the relativistic DD-$\rho\delta$ RMF functional \cite{baran04}  
is included.  
Two Skyrme functionals, SkM$^*$ and the more recent Skyrme-Lyon force  
SkLya represent non-relativistic models.   
The left panel zooms the low density region while the right panel  
shows the high density behavior of $ E_{\rm sym}$. 

The low density part of the symmetry energy is in the meantime 
relatively well constraint by data. Recent NSCL-MSU heavy ion 
data in combination with 
transport calculations are consistent with a value of 
$E_{\rm sym} \approx 31 $ MeV at $\rho_0$ and 
rule out extremely "stiff" and "soft" density 
dependences of the symmetry energy \cite{shetty05}.  
The same value has been extracted \cite{Khoa06} 
from low energy elastic and (p,n) charge exchange 
reactions on isobaric analog states p($^{6}He,^{6}Li^*$)n measured at the HMI. 
At sub-normal densities  recent data points have been 
extracted from the isoscaling behavior of fragment formation 
in low-energy heavy ion reactions with the corresponding experiments 
carried out at Texas A\&M and NSCL-MSU \cite{shetty07}.

However, theoretical extrapolations to supra-normal densities  
diverge dramatically. This is crucial since the high density behavior  
of  $ E_{\rm sym}$ is essential for the structure and the stability of  
neutron stars. 
The microscopic models show a density dependence which can still  
be considered as {\it asy-stiff}. DBHF \cite{dalen04} is thereby stiffer  
than the variational results of Ref. \cite{akmal98}. The density  
 dependence is generally more complex  than in RMF  theory, in  
particular at high densities where $E_{\rm sym}$ shows a non-linear and  
more pronounced increase.   
Fig. \ref{esym_fig} clearly demonstrates the necessity to better constrain  
the symmetry energy at supra-normal densities with the help of heavy  
ion reactions. 

\section{Constraints from heavy ion reactions} 
\begin{figure}[h]
\centerline{
\includegraphics[width=0.7\textwidth]{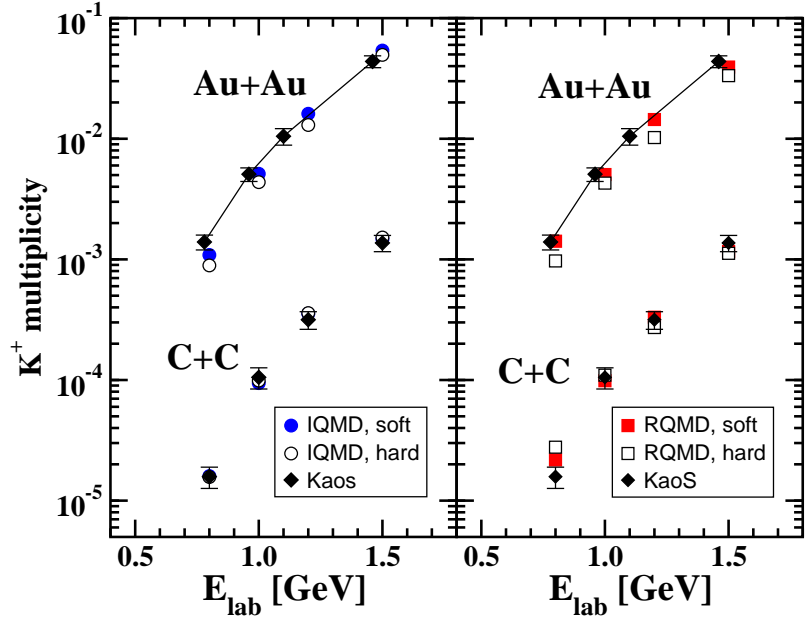}} 
\caption{Excitation function of the $K^+$ multiplicities 
in $Au+Au$ and $C+C$ reactions. RQMD \protect\cite{fuchs01} and 
IQMD \protect\cite{hartnack05} with 
in-medium kaon potential and using a hard/soft nuclear EoS
are compared to data from the KaoS Collaboration \protect\cite{sturm01}. 
The figure is taken from \protect\cite{WCI}.}
\label{fig_ex_2}
\end{figure}
Experimental data which put constraints on the symmetry energy have 
already been shown in Fig. \ref{esym_fig}. The problem of multi-fragmentation 
data from low and intermediate energy reactions is that they are 
restricted to sub-normal densities up to maximally saturation density. 
However, from low energetic isospin diffusion measurements at least the slope of the 
symmetry around saturation density could be extracted \cite{Chen:2004si}. 
This puts already an important constraint on the models when extrapolated to 
higher densities. It is important to notice that the slopes predicted 
by the ab initio approaches (variational, DBHF) shown in Fig. \ref{esym_fig} 
are consistent with the empirical values. Further going attempts to derive 
the symmetry energy at supra-normal densities from particle production 
in relativistic heavy ion reactions \cite{baran04,ferini06,Gaitanos:2003zg} have 
so far not yet led to firm conclusions since the corresponding signals 
are too small, e.g. the isospin dependence of kaon production \cite{Lopez:2007rh}.

Firm conclusions could only be drawn on the symmetric part of the 
nuclear bulk properties.  To explore supra-normal densities one has to 
increase the bombarding energy up to relativistic energies. This was one 
of the major motivation of the SIS100 project at the GSI where - according to 
transport calculation - densities between 
$1\div 3~ \rho_0$ are reached at bombarding energies between $0.1\div 2$ AGeV. 
Sensitive observables are the collective nucleon flow and  
subthreshold $K^+$ meson production. In contrast to the flow signal 
which can be biased by surface effects and the momentum dependence of the
optical potential,  $K^+$ mesons turned out to an excellent probe for the 
high density phase of the reactions. At subthreshold energies the 
necessary energy has to be provided by multiple scattering processes which are 
highly collective effects. This ensures that the majority of the  $K^+$ mesons 
is indeed produced at supra-normal densities. In the following I will concentrate 
on the kaon observable. 
\begin{figure}[h]
\centerline{
\includegraphics[width=0.5\textwidth]{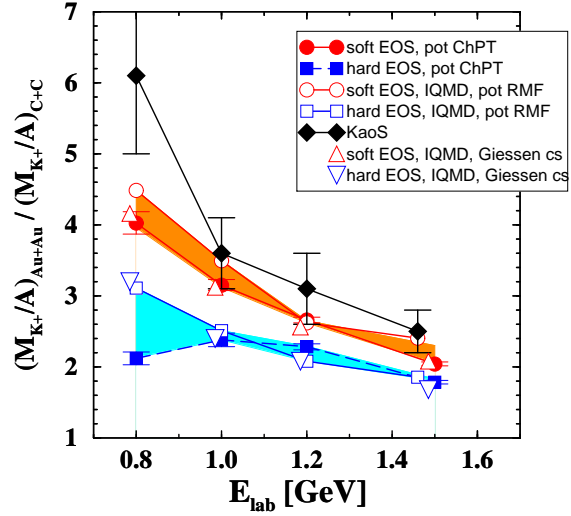}} 
\caption{Excitation function of the ratio $R$ of $K^+$ 
multiplicities obtained in inclusive Au+Au over C+C 
reactions. RQMD \protect\cite{fuchs01} and IQMD 
\protect\cite{hartnack05} calculations are compared 
to KaoS data \protect\cite{sturm01}. Figure is taken from 
\protect\cite{fuchs05}.
}
\label{fig_ratio_1}
\end{figure}

Subthreshold particles are rare probes. However, within the last 
decade the KaoS Collaboration has performed systematic high statistics 
measurements of the $ K^+$ production far below threshold 
\cite{sturm01,schmah05}. Based on this data situation, 
in Ref. \cite{fuchs01} the question 
if valuable information on the nuclear EoS can be extracted 
has been revisited and it has been 
shown that subthreshold  $K^+$ production provides indeed a suitable and 
reliable tool for this purpose.  In subsequent investigations 
the stability of the EoS dependence has been proven \cite{hartnack05,fuchs05}. 

Excitation functions from KaoS \cite{sturm01,kaos99} are 
 shown in Fig. \ref{fig_ex_2} and compared to RQMD \cite{fuchs01,fuchs05} 
and IQMD \cite{hartnack05} calculations. In both cases a soft (K=200 MeV) and 
a hard (K=380 MeV) EoS have been used within the transport approaches. 
Skyrme type forces supplemented by an empirical momentum 
dependence have been used. As expected the EoS dependence 
is more pronounced in the heavy Au+Au system while the light C+C system 
serves as a calibration. 
The effects become even more evident when the ratio $R$ of the 
kaon multiplicities obtained in Au+Au over C+C 
reactions (normalized to the corresponding mass numbers) is built 
\cite{fuchs01,sturm01}. Such a ratio has the advantage that 
possible uncertainties which 
might still exist in the theoretical calculations cancel out 
to large extent. Comparing the ratio shown in Fig. \ref{fig_ratio_1} 
to the experimental data from KaoS \cite{sturm01}, where the increase of $R$ is even more 
pronounced, strongly favors a soft equation of state. 
This result is in agreement with the conclusion drawn from the 
alternative flow observable \cite{dani00,gaitanos01,andronic99,stoicea04}.

\section{Constraints from neutron stars} 
\begin{figure}[h]
\centerline{
\includegraphics[width=0.7\textwidth]{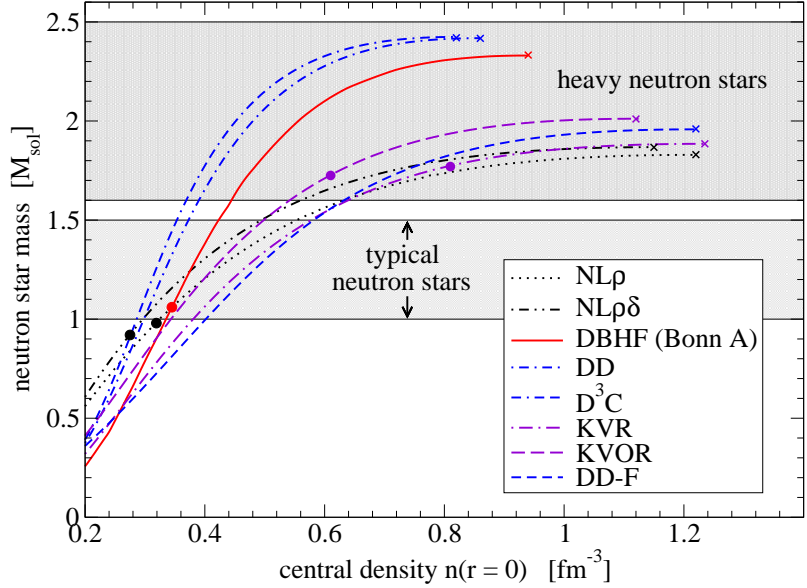}} 
\caption{Mass versus central density for compact star configurations obtained 
for various relativistic hadronic EsoS. 
Crosses denote the maximum mass configurations, filled dots mark the
critical mass and central density values where the DU cooling process becomes
possible. According to the DU constraint, it should not occur in ``typical
NSs'' for which masses are expected from population synthesis
\cite{Popov:2004ey} to lie in the {lower} grey horizontal band.
The dark grey horizontal bands around indicate an conservative estimate for  
the possible range of maximal neutron star masses derived from recent 
observation.
}
\label{mass_fig}
\end{figure}
Measurements of ``extreme'' values,
like large masses or radii, huge luminosities etc.\
as provided by compact stars offer good opportunities to gain deeper insight
into the physics
of matter under extreme conditions. 
There has been substantial progress in recent time from the astrophysical 
side. 

The most spectacular observation was probably the 
recent measurement \cite{NiSp05} on PSR J0751+1807, a millisecond pulsar in a binary 
system with a helium white dwarf secondary, which implied 
a pulsar mass of $2.1\pm0.2\left(^{+0.4}_{-0.5}\right) {\rm M_\odot}$
with $1\sigma$ ($2\sigma$) confidence. This measurement has, however, 
 been revisited and 
corrected down to $M=1.26~M_\odot$ \cite{Nice07}.

There exist, however, several alternative 
observations pointing towards large masses, e.g. the large radius of $R > 12$ km for the 
isolated neutron star RX J1856.5-3754 (shorthand: RX J1856) 
\cite{Trumper:2003we}. 
Measurements of high masses are also reported for compact stars in low-mass
X-ray binaries (LMXBs) as $M=2.0\pm 0.1~M_\odot$ for the compact object in 4U 1636-536
\cite{Barret:2005wd}.    
For another LMXB, EXO 0748-676, 
constraints for the mass $M\ge 2.10\pm 0.28~M_\odot$
{\it and} the radius $R \ge 13.8 \pm 0.18$ km  for the same object 
have been reported \cite{Ozel:2006km}. Very recently even an extremely high 
mass value of  $M= 2.74\pm 0.21~M_\odot$ $(1\sigma$) has been reported for 
the millisecond pulsar PSR J1748-2021B \cite{Freire:2007jd}. According to the authors 
of Ref. \cite{Freire:2007jd} there exists only a  1 \% probability that the pulsar mass 
is below 2 solar masses, and a 0.10 \% probability that it lies within the range 
of conventional neutron stars, i.e. between 1.20 and 1.44 solar masses. Such an
 anomalously large mass would of course strongly constrain 
the equation of state for dense matter, even excluding many-body approaches 
which reach maximum masses around $M=2.3~M_\odot$. 

In Ref. \cite{klaehn06} we applied  more conservative upper and lower 
limits for the maximum mass of $1.6-2.5~M_\odot$ (initiated by the originally reported 
$2\sigma$ range of PSR J0751+1807 \cite{NiSp05}). 
However, even such a weaker condition limits the 
softness of the EoS in neutron star (NS) matter considerably. The corresponding 
figure, Fig. \ref{mass_fig}, shows the 
mass versus central density for compact star configurations obtained by
solving the TOV equations for a compilation of different hadronic EsoS. 
These are relativistic mean field models and the microscopic DBHF model. 
For details see \cite{klaehn06}. Crosses denote the maximum mass 
configurations, filled dots mark the
critical mass and central density values where the DU cooling process becomes
possible.

\begin{figure}[h]
\centerline{
\includegraphics[width=0.6\textwidth]{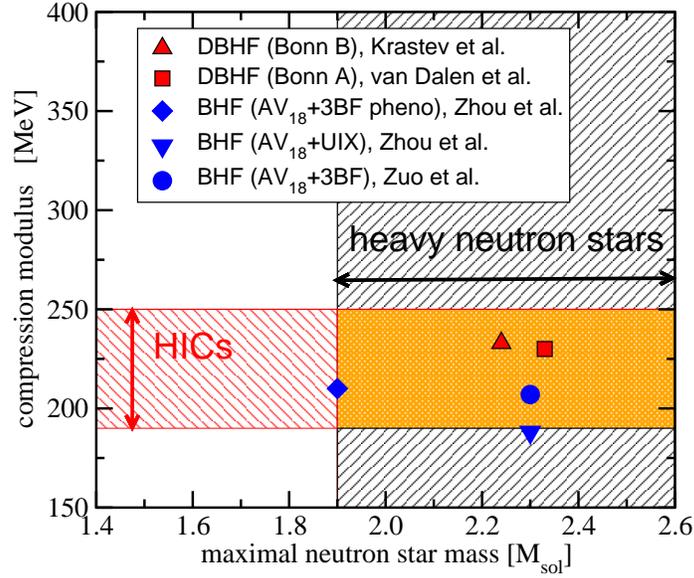}} 
\caption{Combination of the constraints on the EoS derived from the 
maximal neutron star mass criterium and the heavy ion collisions 
constraining the compression modulus. Values of various microscopic 
BHF and DBHF many-body calculations are show.
}
\label{constraint_fig}
\end{figure}

One might now  be worried about an apparent contradiction between the 
constraints derived from neutron stars and those from heavy ion 
reactions. While heavy ion reactions favor a soft EoS, high neutron star masses 
require a stiff EoS. The 
corresponding constraints are, however, 
complementary rather than contradictory. 
Intermediate energy heavy-ion reactions, e.g. subthreshold kaon production, 
constrains the EoS at densities up to $2\div3~\rho_0$ while the maximum NS 
mass is more sensitive to the high density behavior of the EoS. Combining the 
two constraints implies that the EoS should be {\it soft at moderate 
densities and  stiff at high densities}. Such a behavior is predicted 
by microscopic many-body calculations (see Fig.~\ref{constraint_fig}). DBHF, BHF 
or variational calculations, typically, lead to maximum NS masses between  
$2.1\div 2.3~M_\odot$ and are therefore in accordance with most of the high 
mass neutron star measurement (except of masses around or 
above $2.4\div2.5~M_\odot$.) 

This fact is illustrated in Fig. \ref{constraint_fig} which 
combines the results from heavy ion collisions and the maximal mass constraint. 
The figure shows the compression moduli for the EsoS in symmetric nuclear 
matter as well as the maximum neutron star mass obtained with the corresponding 
models, i.e. non-relativistic Brueckner (BHF) as well  as for relativistic 
Brueckner (DBHF) calculations \cite{dalen04,krastev06}. The BHF calculations differ 
essentially in the usage of different 3-body-forces (3-BFs). In particular the 
isospin dependence of 3-BFs is not yet well constrained by nuclear data which 
is reflected in the maximum masses obtained, not so much in the compression moduli. 
The DBHF calculations differ in the elementary NN interaction applied. However, 
here the results for both, compression moduli and 
maximum neutron star masses are rather stable.

Besides the maximum masses there exist several other constraints on the nuclear EoS which can 
be derived from observations of  compact stars, see e.g. 
Refs. \cite{klaehn06,steiner05}. 
Among these, the most promising one is the Direct Urca (DU) process which 
is essentially driven by the proton fraction inside the NS \cite{lattimer91}. 
DU processes, e.g. the neutron $\beta$-decay $n\to p+e^-+\bar\nu_e$,
are very efficient regarding their neutrino production,
even in super-fluid NM and cool NSs too fast to be in accordance
with data from thermally observable NSs.
Therefore, one can suppose that
no DU processes should occur
below the upper mass limit
for ``typical'' NSs, i.e.
$M_{DU}\geq 1.5~M_\odot$ ($1.35~M_\odot$ in a weak interpretation).
These limits come from a population synthesis of young,
nearby NSs~\cite{Popov:2004ey} and masses of NS binaries  ~\cite{NiSp05}. 
While the present DBHF EoS leads to too fast neutrino cooling this 
behavior can be avoided if a phase transition to quark matter is 
assumed \cite{klaehn06b}. Thus a quark phase is not ruled out by the 
maximum NS mass. However, corresponding quark EsoS have to be 
almost as stiff as typical hadronic EsoS \cite{klaehn06b}. 
\section{Summary} 
Heavy ion reactions provide in the meantime reliable constraints on the 
isospin dependence of the nuclear EoS at sub-normal densities up to 
saturation density and for the symmetric part up to - as an conservative estimate - 
two times saturation density. These are complemented by astrophysical 
constraints derived from the measurements of extreme values for 
neutron star masses. As long as the neutron star mass is below $2.3~M_\odot$, both, 
the heavy ion constraint as well as the astrophysical constraint 
is in fair agreement with 
the predictions from nuclear many-body theory. If, however, a maximum mass 
around or above  $2.5~M_\odot$ \cite{Freire:2007jd} will be established, this requires an 
extremely stiff EoS which demands for new physical pictures.


\end{document}